\documentclass[twocolumn,aps,prb,showpacs,floatfix,superscriptaddress]{revtex4}

\usepackage{epsfig}

\newcommand{\beq}{\begin{equation}}
\newcommand{\eeq}{\end{equation}}
\newcommand{\bea}{\begin{eqnarray}}
\newcommand{\eea}{\end{eqnarray}}

\begin{document}

\title{Local moment fluctuations in an optimally-doped high T$_c$ superconductor}
\author{D. Reznik}
\affiliation{Forschungszentrum Karlsruhe, Institut f\"ur
Festk\"orperphysik, Postfach 3640, D-76021 Karlsruhe, Germany}
\affiliation{Laboratoire Leon Brillouin, C.E.A./C.N.R.S.,
F-91191-Gif-sur-Yvette CEDEX, France}
\author{J.-P. Ismer}
\affiliation{Max-Planck-Institut f\"ur Physik Komplexer Systeme,
D-01187 Dresden, Germany} \affiliation{Institute f\"ur
Mathematische und Theoretische Physik, TU-Braunschweig, 38106
Braunschweig, Germany}
\author{I. Eremin}
\affiliation{Max-Planck-Institut f\"ur Physik Komplexer Systeme,
D-01187 Dresden, Germany} \affiliation{Institute f\"ur
Mathematische und Theoretische Physik, TU-Braunschweig, 38106
Braunschweig, Germany}
\author{L. Pintschovius}
\affiliation{Forschungszentrum Karlsruhe, Institut f\"ur
Festk\"orperphysik, Postfach 3640, D-76021 Karlsruhe,
Germany}
\author{T. Wolf}
\affiliation{Forschungszentrum Karlsruhe, Institut f\"ur
Festk\"orperphysik, Postfach 3640, D-76021 Karlsruhe, Germany}
\author{M. Arai}
 \affiliation{Institute of Materials
Structure Science, KEK, Tsukuba 305-0801, Japan}
\author{Y. Endoh}
\affiliation{Synchrotron Radiation Research Center, Japan Atomic
Energy Research Institute, Hyogo 679-5148, Japan}
\author{T. Masui}
\author{S. Tajima}
\affiliation{Department of Physics, Osaka University, Toyonaka,
Osaka 560-0043, Japan}

\date{\today}

\begin{abstract}
We present results of neutron scattering experiments on
YBa$_2$Cu$_3$O$_{6.95}$ (T$_c$=93K). Our results indicate that
magnetic collective modes due to correlated local moments are
present both above and below T$_c$ in optimally doped YBCO. The
magnon-like modes are robust and not overdamped by itinerant
particle-hole excitations, which may point at a substantial
static or slowly fluctuating charge inhomogeneity. We compare the
experimental results to predictions of the Fermi liquid (FL)
theory in the Random Phase Approximation (RPA).
\end{abstract}

 \maketitle

High temperature superconductivity occurs in layered copper oxides
at compositions between undoped insulating and strongly doped
conventional metallic phases. Some propose that exotic states
inherited from the antiferromagnetic parent Mott insulators are
essential for superconductivity \cite{1} whereas others claim
that superconductors with the highest transition temperatures,
T$_c$, conform to the standard model of metals, the Fermi liquid
theory\cite{2}.

Magnetic spectra of the copper oxides change dramatically as a
function of doping. At $x=0$, these compounds are Mott insulators
with strong on-site repulsion that localizes electrons whose spins
order antiferromagnetically. Their low energy excitations are spin
waves (magnons), which disperse upward from the antiferromagnetic
(AF) ordering vector\cite{3,4}. Doping disrupts long range AF
order, but dispersive excitations similar to magnons remain in
underdoped superconductors ($0<x<.13$) at high
energies\cite{5,6,7,8}. However, such magnetic excitations have
not so far been detected in superconductors with T$_c>80$K.
Previous experimental studies of the compounds with the highest
T$_c$s focused primarily on the resonance feature around
40meV\cite{9,10,11} appearing only below T$_c$\cite{12}. Both
Fermi liquid and non-Fermi liquid scenarios have been proposed to
explain the resonance\cite{1,2,5,13,14,15,16,17}.

Here, we focus on magnetic excitations in optimally doped
YBa$_2$Cu$_3$O$_{6.95}$ (T$_c$ = 93 K) above T$_c$.   Since
previous studies failed to clearly distinguish\cite{12,18} the
magnetic signal from the nuclear background, we used much longer
counting times (more than 1hr/point in some scans) to
significantly reduce statistical error. We report that the
collective magnetic excitations do exist in
YBa$_2$Cu$_3$O$_{6.95}$ and that their intensity is appreciable
in comparison to the magnon intensity in the parent insulator and
the underdoped YBa$_2$Cu$_3$O$_{6.6}$.

The experiments were performed on the 1T triple-axis spectrometer at
the ORPHEE reactor at Saclay utilizing a double focusing Cu111
monochromator and a pyrolytic graphite (PG002) analyzer. The high
quality sample of YBa$_2$Cu$_3$O$_{6.95}$ was the same as in an
earlier study\cite{19}.

\begin{figure}[h]
\includegraphics[width=7cm]{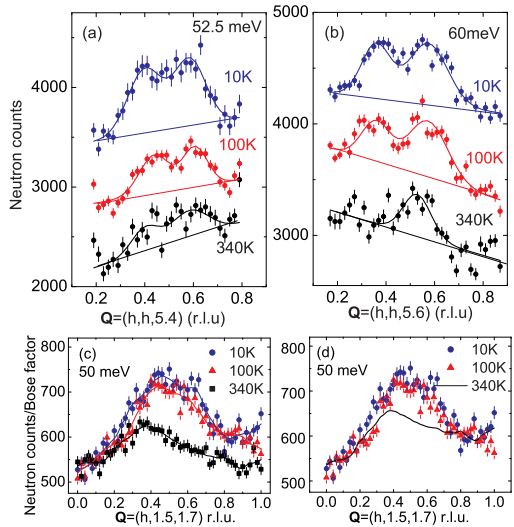}
\caption{(Color online) Raw data and background subtraction
procedure. (a,b) Scans at 52.5 and 60meV at 10K, 100K, and 340K
offset by a constant for clarity. Lines are guides to for the
eye. Note that the peak in (b) at 340K is most certainly
nonmagnetic. (c) Scans at 50meV through the antiferromagnetic
wave vector divided by the Bose factor. Solid lines represent
smoothed data. The feature centered at $h$=0.5 at 10K and 100K is
magnetic because it is suppressed at 330K as the underlying
correlations become weaker, leaving behind a "hump" of nuclear
scattering peaked at $h$=0.4. The background levels near $h$=0
and $h$=1 do not exactly match due to small nuclear contributions
whose temperature-dependence does not follow the Bose factor
(e.g. multi-phonon or incoherent elastic nuclear scattering).
Linear corrections were added to the 100K and 330K data to match
the backgrounds near $h$=0 and $h$=1 with the result shown in
(b). These corrections were small is all cases (the 50 meV data
are the worst case); (d) 50 meV scans after adding the linear
terms to the 100K and 330K data. We assign the intensity
difference between 10K/100K data and 330K curve to magnetic
scattering. Error bars represent s.d.} \label{fig1}
\end{figure}

Following Ref. \onlinecite{20}, magnetic collective modes were
separated from the nuclear background by subtracting spectra
recorded at a high temperature where the magnetic modes are
largely suppressed. Fig. \ref{fig1}(a) and (b) give examples of
the raw data at 52.5 and 60 meV where there is a clear double
peak structure around $h$=0.5, the AF wavevector of the undoped
parent compounds, at 10K and 100K. The intensity of this feature
is reduced on heating above 300K, as expected from magnetic
collective modes. Some or all intensity that remains at the high
temperatures may be magnetic but to be on the conservative side,
we subtract the entire high T spectrum divided by the Bose factor
from the 10K and 100K data.

\begin{figure}[h]
\includegraphics[width=7cm]{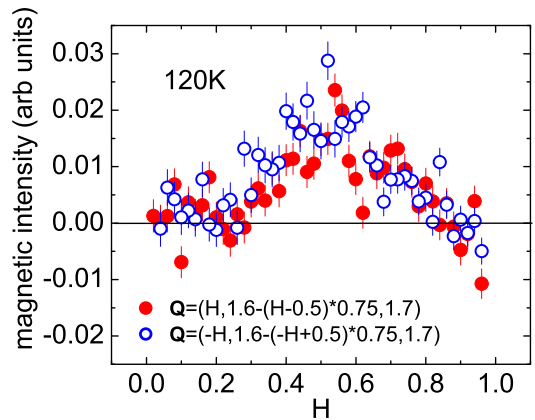}
\caption{(Color online) Magnetic intensity at 120K at 55meV
extracted by the procedure described in the text and Fig.
\protect\ref{fig1}. The data in the vicinity of equivalent AF
wave vectors {\bf Q}=(1.5, 0.5, 1.7) (left panel) and {\bf
Q}=(1.5,-0.5,1.7) (right panel) are shown in arbitrary units but
on the same scale. Small differences between the two curves are
consistent with different resolution functions at the two
wavevectors and/or statistical error. This figure demonstrates
that the intensity that we assign to magnetic scattering has the
correct symmetry.} \label{fig2}
\end{figure}

The measured energies were chosen to minimize systematic errors
due to contamination by nuclear scattering. The biggest error in
our experiment is an underestimate of the magnetic signal because
some of it (10-20$\%$) may remain at high temperatures where the
background was measured. Any broad continuum is indistinguishable
from the background and is also not included in the extracted
intensities. Fig. \ref{fig2} shows that normal state magnetic
intensity extracted by this procedure obeys crystal symmetry
appearing with a similar intensity in the vicinity of equivalent
AF wavevectors ${\bf Q}=(1.5, 0.5, 1.7)$ and $(1.5,-0.5,1.7)$.
The observed magnetic signal was centered at equivalent
antiferromagnetic wavevectors ${\bf Q}=(1.5, 0.5, 1.7)$, ${\bf
Q}=(0.5, 0.5, 5.4)$, and ${\bf Q} = (1.5, 1.5, 1.7)$ in units of
$(2\pi /a, 2 \pi/b, 2 \pi /c)$, (where $a$, $b$, and $c$ are
lattice constants (data not shown for the latter)). Its
intensities in different Brillouin zones agree with the magnetic
form factor confirming the validity of this procedure. Since YBCO
is a bi-layer compound, it has magnetic excitations of acoustic
and optic character. Our choice of L=1.7 and 5.4 selects acoustic
magnetic excitations; a separate study will be necessary to
investigate optic modes.
\begin{figure}[h]
\includegraphics[width=7cm, clip]{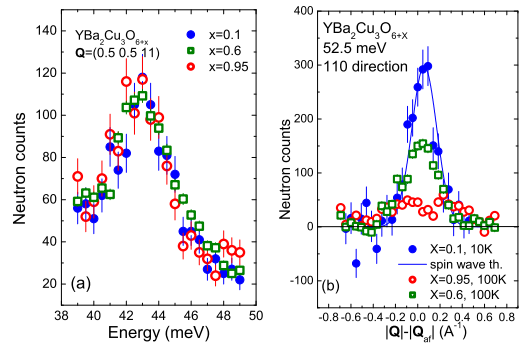}
\caption{(Color online) Comparison of magnetic scattering of the
YBCO$_{6.1}$ YBCO$_{6.6}$, and YBCO$_{6.95}$ samples.(a) Spectra
of the Cu-O buckling mode scaled by the factor equal to the ratio
of the volumes of the three samples. (b) Magnetic scattering
intensities plotted on the same intensity scale after
normalization by the volume ratio extracted from comparison of
phonon intensities in (a).}\label{fig4}
\end{figure}
\begin{figure}[h]
\includegraphics[width=7cm]{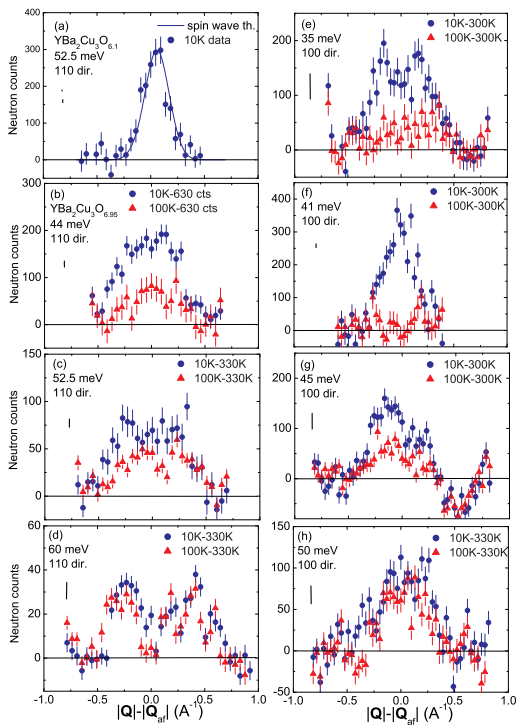}
\caption{(Color online) Magnetic intensities at different
energies. (a) Magnon peak in YBa$_2$Cu$_3$O$_{6.1}$ at 52.5 meV
after subtraction of a flat background. Solid line represents the
spin wave model in Ref. \cite{6} convoluted with the spectrometer
resolution. (b-h) Magnetic scattering in YBa$_2$Cu$_3$O$_{6.95}$
as a function of the distance from the antiferromagnetic ordering
wave vector. All spectra have been normalized to the same scale.
A constant background of 530 and 630 counts was subtracted in (a)
and (b) respectively, whereas the other magnetic spectra were
obtained as described in the text and Fig.\protect\ref{fig1}. The
weak broad 100K signal at 35meV may be an artifact of the
background subtraction procedure. Error bars represent s.d. Our
estimate of systemetic errors given by vertical black lines is
based on the temperature variation of the background.}
\label{fig3new}
\end{figure}

Fig. \ref{fig4} shows the magnon scattering intensity of
insulating YBa$_2$Cu$_3$O$_{6.1}$ and of the underdoped Tc=60K
superconductor, YBa$_2$Cu$_3$O$_{6.6}$, at 52.5meV plotted
together with that of the optimally-doped sample at 100K on the
same intensity scale. IN the underdoped samples it was obtained
by subtracting a constant background since significant magnetic
intensity is still present at 300K. To correct for sample volume
we scaled together spectra of the bond-buckling phonon at E = 42.5
meV. Its eigenvector is independent of doping because it does not
contain any chain oxygen vibrations\cite{12}. The phonon was
measured in identical spectrometer configurations (at {\bf Q}=
(0.5, 0.5, 11)) in the two samples (Fig. \ref{fig4}a). Therefore,
no resolution corrections were needed. The magnon peak intensity
in YBa$_2$Cu$_3$O$_{6.1}$ is six times stronger than that of the
normal state magnetic signal in YBa$_2$Cu$_3$O$_{6.95}$ whereas
the peak intensity of the T$_c$=60K sample measured at 100K was
three times stronger. However, the magnetic signal in
YBa$_2$Cu$_3$O$_{6.95}$ is significantly broader than in the
insulating phase and in the underdoped sample (In the insulator
it is nearly {\bf Q}-resolution-limited). Thus their {\bf
q}-integrated intensities of the three samples must be of the
same order of magnitude around 50meV  at 100K after
two-dimensional {\bf Q}-integration. In making this statement, we
are assuming that magnetic scattering is broad in {\bf q} and has
a fourfold symmetry around the reduced antiferromagnetic wave
vector {\bf Q}$_{AF}$=(0.5,0.5) in a twinned sample like ours.

\begin{figure}[h]
\includegraphics[width=7cm]{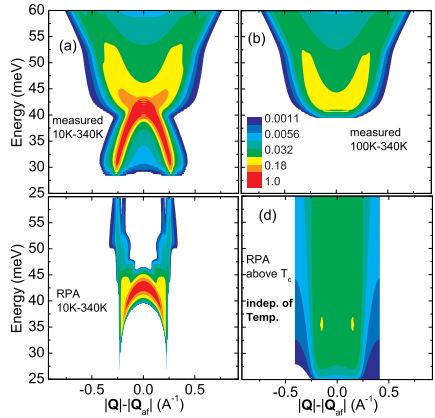}
\caption{(Color online) Magnetic susceptibility of
YBa$_2$Cu$_3$O$_{6.95}$. (a/b) Experimentally measured magnetic
susceptibility difference between 10K/100K respectively and 300K
in the $1 1 0$ direction after removing effects of the
resolution. Different ad. hoc. functional forms were tried until
their convolution with the spectrometer resolution agreed with
the data. (c) Calculations based on FL/RPA (Ref.
\protect\cite{15}) for the difference between the superconducting
state (10K) and normal state (300K). (d) RPA result for the
normal state. Here the difference between 100K and 300K is
negligibly small (in contrast with experiment), so only 100K
result is shown without taking the difference. Note the
logarithmic intensity scale when comparing experimental and
calculated spectra.} \label{fig5}
\end{figure}

Fig. \ref{fig3new} (b-h) shows magnetic spectra of
YBa$_2$Cu$_3$O$_{6.95}$ measured at different energies at 10K and
100K, respectively, and Fig. \ref{fig5}(a,b) shows the cut along
$[1 1 0]$ after removing effects of the resolution. It is based on
the data shown in Fig. \ref{fig3new} and Ref. \onlinecite{19}. At
100K we observe a branch above 43 meV dispersing away from ${\bf
Q}_{AF}$. Its {\bf q}-width is 0.3$\AA^{-1}$ full width at half
maximum (FWHM), which corresponds to a correlation length of about
20$\AA$. Upon cooling to 10K, the scattering intensity below 60
meV increases and a downward-dispersing branch with a constant
{\bf q}-integrated intensity appears between 33 and 41 meV below
T$_c$ as shown in Refs. \onlinecite{20,19}. The {\bf q}-width of
the lower branch is relatively narrow (Fig. 5(a)), corresponding
to coherent domains of about 55$\AA$\cite{19}. The
upward-dispersing branch is significantly less steep than the
magnon dispersion in insulating YBa$_2$Cu$_3$O$_{6.1}$. (Fig.
5a,b) Assuming linear extrapolation, it crosses the zone boundary
around 130/100 meV in the $[1 0 0] /[1 1 0]$ direction,
respectively, compared with 220meV for the magnons in the
insulator\cite{4}. It also disperses significantly less steeply
than the upper branch in YBa$_2$Cu$_3$O$_{6.6}$, which has a much
narrower lineshape at 52.5meV (Fig. 3) with the magnetic
intensity remaining near to {\bf q}=(0.5,0.5).\cite{7}.

Fig. \ref{fig5}(c,d) shows the prediction of the FL theory in the
standard RPA approximation (FL/RPA)\cite{13,14,15,16} using the
calculation based on Ref.\cite{16}. The RPA expression for the
spin susceptibility is:
\begin{equation}
\chi^{+-}_{RPA}({\bf q}, \omega)=\frac{\chi^{+-}_{0}({\bf q},
\omega)}{1-g_{\bf q}\chi^{+-}_{0}({\bf q}, \omega)}
\end{equation}
where $\chi^{+-}_{0}({\bf q}, \omega)$ is the bare transverse
susceptibility and $g_{\bf q}$ is an electron-electron
interaction (four-point vertex), which in general can be momentum
dependent\cite{13,14,15,16}. Once the parameters of the model were
picked to reproduce the known Fermi surface, band width of
optimally-doped copper oxides, the superconducting gap\cite{22},
and the resonance peak energy, there was no further possibility to
adjust parameters to alter the calculation results in any
significant way. Imaginary part of $\chi^{+-}_{RPA}({\bf q},
\omega)$ is proportional to the neutron scattering cross section
and can be directly compared with experiment. Below T$_c$ sharp
downward-dispersing resonance collective modes appear in the
calculation due to an excitonic effect inside the d$_{x^2-y^2}$
-wave superconducting gap (Fig. 5c)\cite{16}. This feature
appears to be in good agreement with the experiment (Fig. 5 a,c).
The calculation also predicts an upward-dispersing feature below
T$_c$. However it is an order of magnitude weaker (relative to the
downward-dispersing branch) and has a steeper effective
dispersion then the measured magnetic signal.

Above T$_c$ FL/RPA predicts only a broad temperature-independent
particle-hole continuum and no collective modes (Fig. 5d). Within
a simple Fermi-liquid picture one expects weak temperature
dependence of the Im$\chi$ proportional to the $k_BT/E_F$ with
$E_F$ being the Fermi energy. In optimally-doped YBCO this ratio
is too small to be detected at the energies that we probed. In
contrast with the FL/RPA result, our experiment shows that the
spin signal grows significantly on cooling from 340K to 100K.
Including the renormalization of the quasiparticles due to
scattering by the spin fluctuations within a so-called
fluctuation exchange approximation (FLEX) yields a much stronger
temperature dependence of the spin excitations at low energies
but not at the high energies that we investigated \cite{dahm}. It
is not entirely clear how further renormalization of the
interaction strength (vertex corrections) will affect the results
of the FLEX calculation. Our preliminary estimate is that it will
further reduce the predicted temperature
dependence.

The above analysis leads us to conclude that the FL/RPA picture does
explain magnetic intensity above {$\sim$45meV} in neither the normal
state nor in the superconducting state. The 52.5 and 60 meV data
(Figs. 1, 3c,d) show incommensurate collective modes both above and
below T$_c$ whose dispersion is similar to magnons in the parent
insulator. (Fig. 5b) (Note that the scans at 55meV in Fig. 2 do not
have a double-peak structure because they were centered not at
{\bf$_Q$}=(1.5 0.5 1.7), but at one of the incommensurate
satellites: {\bf$ Q$}=(1.6 0.5 1.7).) Based on this observation we
assign these modes to collective excitations of local moments
inherited from the insulating
phase\cite{batista,eremin,sega,sherman} as opposed to itinerant
quasiparticles inherited from the overdoped conventional FL phase.
These local moments must coexist with the itinerant particle-hole
continuum spanned by the Fermi surface.

Our results suggest a qualitative similarity between the YBCO
family and the La$_{2-x}$Sr$_x$CuO$_4$ family of high T$_c$
superconductors. Wakimoto et al. \cite{27a} found that low energy
magnetic excitations peaked near {\bf Q}$_{AF}$ disappear upon
increased doping whereas the high energy fluctuations persist far
into the overdoped regime\cite{27b}. In YBa$_2$Cu$_3$O$_{6+x}$
the lowest energy for detecting magnetic excitations due to local
moments also increases with increasing doping reaching
approximately 45 meV at optimal doping as shown in our
investigation. This apparent gap in the measurable magnetic signal
probably appears because the magnetic modes are not magnons in
the strict sense but rather precursor fluctuations to the
formation of antiferromagnetic clusters. Developing a detailed
theory of this phenomenon is outside the scope of our study.

One possibility that can be ruled out is that our sample may
still be in the pseudogap state at 100K and thus the observed
local magnetism is associated with the pseudogap. Figure 2 shows
that the magnetic signal persist at least up to 120K, which is
above the pseudogap temperature reported for T$_c$=93K YBCO. Thus
the magnetic signal above T$_c$ at optimal doping is not
associated with the pseudogap.\cite{26a}

Acknowledgement: D.R. would like to thank S.A. Kivelson, J.
Zaanen, and M. Vojta for valuable comments on earlier versions of
the manuscript. I.E. would like to thank T. Dahm for helpful
discussions.

\end{document}